\documentclass[a4paper,aps,prd,10pt,preprintnumbers,showpacs,superscriptaddress,nofootinbib,amsmath,amssymb,floatfix
,twocolumn
]{revtex4-1}
\def\JCAPstyle#1{}
\usepackage{enumitem}
\usepackage{graphicx}
\usepackage[utf8]{inputenc}
\usepackage[T1]{fontenc}
\usepackage{cmap}
\usepackage{graphicx}
\usepackage{dcolumn}
\usepackage{bm}
\usepackage{bbm}
\usepackage{amsfonts}
\usepackage{amsmath}
\usepackage{enumitem}
\usepackage{lipsum}
\usepackage{graphicx}
\usepackage[utf8]{inputenc}
\usepackage[bottom]{footmisc}
\usepackage{cmap}
\usepackage{subcaption}
\usepackage{braket}
\usepackage{braket}
\usepackage[table]{xcolor}
\usepackage{amsmath}
\usepackage{cleveref}
\usepackage{fmtcount}
\usepackage{dblfnote}
\usepackage{float}
\usepackage{subfig}
\usepackage{subcaption}
\usepackage{amssymb}
\usepackage{natbib}
\usepackage{blindtext}

\begin{document}
\title{A New Approach for Calculation of Quasi-Normal Modes and Topological Charges of Regular Black Holes}
\author{MY Zhang}
\email{myzhang94@yeah.net}
\affiliation{College of Computer and Information Engineering, Guizhou University of Commerce, Guiyang, 550014, China.}
\author{F Hosseinifar}
\email{f.hoseinifar94@gmail.com}
\affiliation{Department of Physics, Faculty of Science, University of Hradec Kr\'{a}lov\'{e}, Rokitansk\'{e}ho 62, 500 03 Hradec   Kr\'{a}lov\'{e}, Czechia}
\author{H Chen}
\email{haochen1249@yeah.net}
\affiliation{School of Physics and Electronic Science, Zunyi Normal University, Zunyi 563006,PR China
}
\author{T Sathiyaraj}
\email{sathiyaraj133@gmail.com}
\affiliation{Institute of Actuarial Science and Data Analytics, UCSI University, Kuala Lumpur 56000, Malaysia}
\author{H Hassanabadi}
\email{hha1349@gmail.com}
\affiliation{Department of Physics, Faculty of Science, University of Hradec Kr\'{a}lov\'{e}, Rokitansk\'{e}ho 62, 500 03 Hradec   Kr\'{a}lov\'{e}, Czechia}
\affiliation{Institute of Actuarial Science and Data Analytics, UCSI University, Kuala Lumpur 56000, Malaysia}
\begin{abstract}
This study examines the properties of a special regular black hole. This analysis investigates the Hawking temperature, remnant radius and mass, as well as the effect of parameter $\xi$ on thermodynamic quantities like entropy, heat capacity, and free energy. The emission rate, evaporation process, quasi-normal modes by calculating Rosen-Morse potential, and topological behavior of the black hole are also explored.
\\\textit{Keywords}: Regular black holes, topological charge, thermodynamic properties
\end{abstract}
\maketitle
\section{Introduction}\label{Sec1}
The standard cosmological model has been highly successful in describing the large-scale structure and evolution of the universe. However, there are some outstanding issues that are not well explained by this model. Therefore, alternative theories, such as modified gravity have been proposed to address these problems \cite{Clifton,Shankaranarayanan,Capozziello,Bonanno}. These modified gravity theories postulate the existence of additional fields or modifications to the gravitational sector, which can lead to predictions and observational signatures that differ from the standard model \cite{Nojiri,Ashtekar}.\\
The authors in Ref. \cite{Bonanno} presented a theoretical model for the collapse of dust in the concept of asymptotically safe gravity. They introduced a function that is derived from the Reuter fixed point in asymptotically safe gravity and investigated the coupling of the matter Lagrangian and the gravitational sector. In this way, the gravitational coupling vanishes at high energies, preventing the formation of singularities during the collapse of dust. The introduced metric is given by
\begin{align}\label{ds2}
ds^2=-f(r)dt^2+\frac{1}{f(r)}dr^2+h(r) \left(d \theta^2+\sin ^2 \theta d \phi^2\right),
\end{align}
where $f(r)$ and $h(r)$ are given by
\begin{align}
f(r)&=1-\frac{r^2}{3 \xi } \log \left(\frac{6 M \xi }{r^3}+1\right),\label{fr}\\
h(r)&=r^2.\label{hr}
\end{align}
When parameter $\xi$ tends to zero, metric reduces to the Schwarzschild black hole.\\
In Ref. \cite{Stashko}, the author investigated some features of this metric such as shadow radius, quasi-normal modes by Chebyshev pseudo-spectral and Eikonal approaches, and greybody bounds.
\\In this work, we first provide an introduction, and then in Sec. \ref{Sec2}, we investigate the Hawking temperature, remnant mass, and radius. In Sec. \ref{Sec3}, we calculate some thermodynamic properties of the black hole, namely the entropy, heat capacity, and free energy. Subsequently, considering the shadow radius and the remnant radius, in Sec. \ref{Sec5} we calculate the emission rate and the evaporation time of the black hole. We also calculate the quasi-normal modes for the expanded form of the metric using the Rosen-Morse potential in Sec. \ref{Sec8}. In Sec. \ref{Sec6} and \ref{Sec7}, we calculate the topological phase transition for several different vector potentials and their topological charge. Finally, in Sec. \ref{Sec10}, we present the summary and conclusion of the work.
\section{Remnant Radius and Mass}\label{Sec2}
As author discussed in Ref. \cite{Stashko}, three regimes which bounds parameter $\xi$ are existed. Case $\xi\in(0,\,0.4565M^2)$ which consists of two horizon (inner and outer) radii, case $\xi=0.4565M^2$ that two horizon radii coincides to a single horizon and case $\xi>0.4565M^2$ which is horizonless.
In this work we choose the first regime where $0<\xi<0.4565M^2$. At horizon $f(r)=0\big|_{r=r_h}$, thus one can find mass as a function of $r_h$
\begin{eqnarray}
M(r_h)=\frac{r_h^3}{6 \xi } \left(e^{\frac{3 \xi }{r_h^2}}-1\right).\label{Mrh}
\end{eqnarray}
Hawking temperature of the black hole is obtained from
\begin{eqnarray}
T_H = \frac{1}{4\pi}\,\partial f(r)/\partial r\big|_{r=r_h}.\label{T1}
\end{eqnarray}
Replacing mass from Eq, \eqref{Mrh} in Eq. \eqref{T1} yields
\begin{eqnarray}
T_H=\frac{r_h}{4 \pi  \xi } \left(1- e^{-\frac{3 \xi }{r_h^2}}-\frac{2 \xi }{r_h^2}\right).\label{TH}
\end{eqnarray}
\begin{figure}[ht!]
\centering
	\begin{subfigure}[b]{0.5\textwidth}
	\centering
	\includegraphics[scale = 0.48]{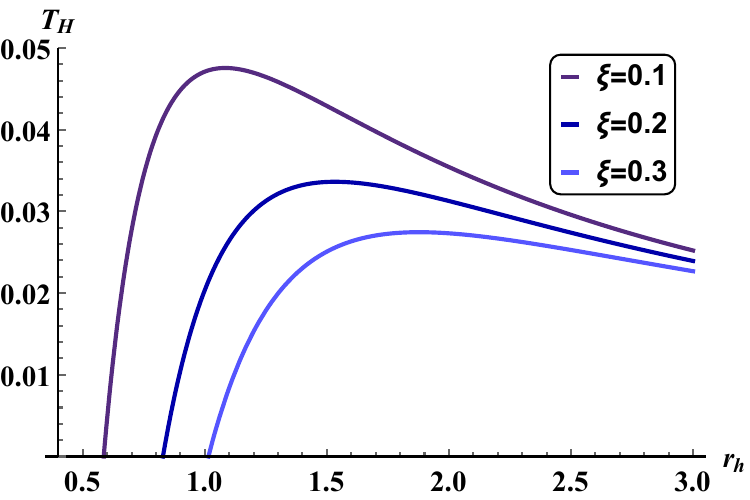} \hspace{-0.2cm}\\
    \end{subfigure}%
    \caption{Hawking temperature curve.}
    \label{fig:TH}
\end{figure}
Remnant radius of the black hole is calculated from $T_H=0\big|_{r=r_{rem}}$ \cite{remnant,remnant2}, and is given by
\begin{eqnarray}
r_{rem}=1.85247 \sqrt{\xi }.\label{rRem}
\end{eqnarray}
Replacing the remnant radius  $r_{rem}$ obtained in Eq. \eqref{rRem}, by the horizon radius in Eq. \eqref{Mrh}, the remnant mass of the black hole is calculated as
\begin{eqnarray}
M_{rem}=1.48012 \sqrt{\xi }.
\end{eqnarray}
\section{Entropy, Heat Capacity, Free Energy}\label{Sec3}
\begin{figure}[ht!]
\centering
	\begin{subfigure}[b]{0.5\textwidth}
	\centering
	\includegraphics[scale = 0.48]{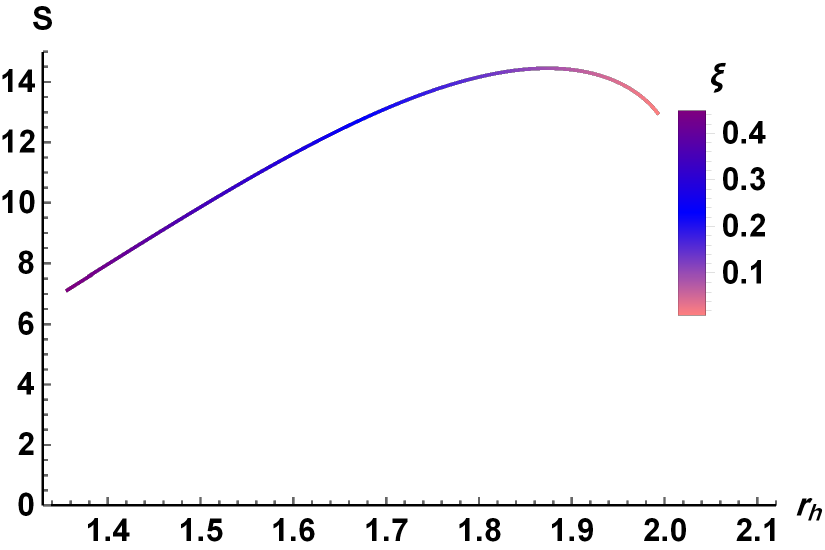} \hspace{-0.2cm}\\
    \end{subfigure}%
    \caption{Entropy versus $r_h$, where $r_h$ is calculated by varying parameter $\xi$ in the range of $(0,\,0.4565)$. Maximum of entropy is located at $r_h=1.780$.}
    \label{fig:entropy}
\end{figure}

\begin{figure}[ht!]
\centering
	\begin{subfigure}[b]{0.5\textwidth}
	\centering
	\includegraphics[scale = 0.48]{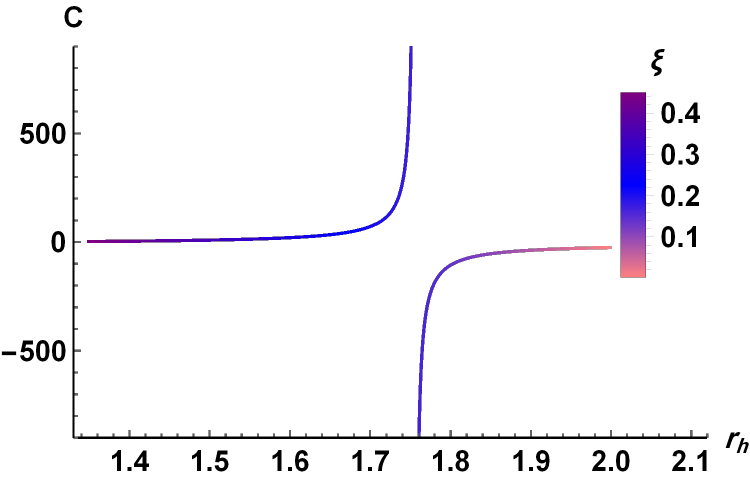} \hspace{-0.2cm}\\
    \end{subfigure}%
    \caption{Heat capacity versus $r_h$, where $r_h$ is calculated by varying parameter $\xi$ in the range of $(0,\,0.4565)$. The value of heat capacity diverges as $r_h=1.755$.}
    \label{fig:HeatCapa}
\end{figure}

\begin{figure}[ht!]
\centering
	\begin{subfigure}[b]{0.5\textwidth}
	\centering
	\includegraphics[scale = 0.48]{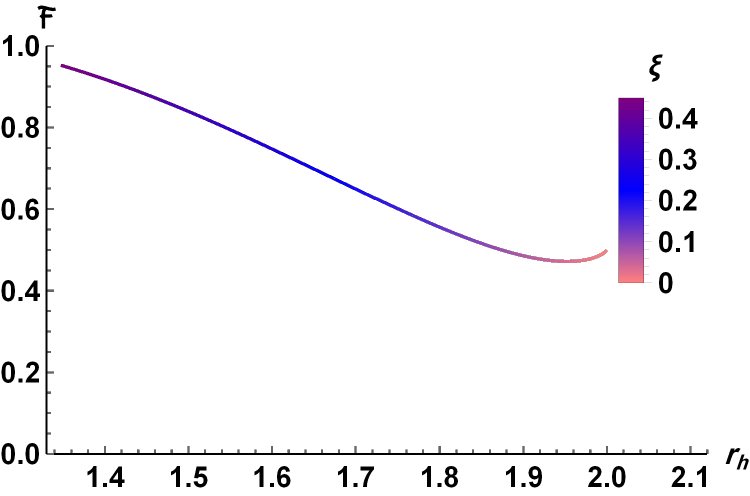} \hspace{-0.2cm}\\
    \end{subfigure}%
    \caption{Helmholtz free energy versus $r_h$, where $r_h$ is calculated by varying parameter $\xi$ in the range of $(0,\,0.4565)$. The minimum value of free energy occurs at $r_h=1.952$.}
    \label{fig:Helm}
\end{figure}

Black hole entropy is given by \cite{remnant}
\begin{align}
\nonumber
S&=\int\frac{\text{d} M}{T_H}
\\&=\pi  \left(r_h^2 e^{\frac{3 \xi }{r_h^2}}-3 \xi  \text{Ei}\left(\frac{3 \xi }{r_h^2}\right)\right),
\end{align}
where Ei represents the exponential integral function. When $\xi$ tends zero, entropy becomes $S=\pi r_h^2$. The illustration of entropy versus parameter $r_h$ is shown in Fig. \ref{fig:entropy}, where $r_h$ is calculated by varying parameter $\xi$.
\\Heat capacity is given by
\begin{align}
\nonumber
C&=\frac{\text{d} M}{\text{d} T_H}\\
&=\frac{2 \pi  r_h^2 \left(e^{\frac{3 \xi }{r_h^2}} \left(r_h^2-2 \xi \right)-r_h^2\right)}{2 \xi -e^{-\frac{3 \xi }{r_h^2}} \left(6 \xi +r_h^2\right)+r_h^2}.
\end{align}
Fig. \ref{fig:HeatCapa} demonstrates heat capacity curve of the black hole.
Evidently, for larger $\xi$, heat capacity becomes positive, at $r_h=1.755$ a phase transition occurs and heat capacity takes a negative sign. As known, a black hole system exhibiting a positive specific heat capacity is indicative of thermodynamic stability, because it necessitates the injection of energy to drive an increase in temperature. Conversely, a negative specific heat capacity referred to thermodynamic instability, where the black hole cools down despite absorbing energy, which indicates its peculiar thermal response.
\\Helmholtz free energy of the system is given by
\begin{align}
F=& M(r_h)- S T_H\label{free}\nonumber
\\ \nonumber
=&
\frac{r_h}{12 \xi } \left(9 \xi  \left(-\frac{2 \xi }{r_h^2}-e^{-\frac{3 \xi }{r_h^2}}+1\right) \text{Ei}\left(\frac{3 \xi }{r_h^2}\right)\right.
\\&\left. -e^{\frac{3 \xi }{r_h^2}} \left(r_h^2-6 \xi \right)+r_h^2\right).
\end{align}
Fig. \ref{fig:Helm} represents the free energy curve as a function of horizon radius.
\section{Emission Rate, Evaporation Time}\label{Sec5}
\begin{figure}[ht!]
\centering
	\begin{subfigure}[b]{0.5\textwidth}
	\centering
	\includegraphics[scale = 0.48]{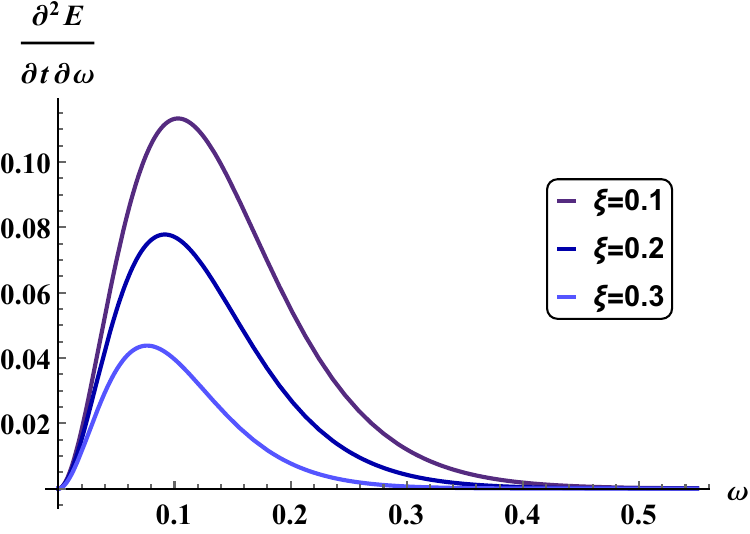} \hspace{-0.2cm}\\
    \end{subfigure}%
    \caption{Emission rate against frequency, where $M=1$.}
    \label{fig:emissionrate}
\end{figure}

Photon radius of the black hole is calculated from \cite{Claudel,Virbhadra}
\begin{eqnarray}
\left(r^2 \frac{\partial f(r)}{\partial r} -2r f(r)\right)\bigg|_{r=r_{ph}}\label{rph}=0,
\end{eqnarray}
and is given by \cite{Stashko}
\begin{align}
r_{ph}=M +2M \sin \left(\frac{1}{3} \cos ^{-1}\left(\frac{3 \xi }{M^2}-1\right)+\frac{\pi }{6}\right).
\end{align}
Shadow radius is given by
\begin{eqnarray}
r_{sh}=\frac{r_{ph}}{\sqrt{f(r_{ph})}} = \frac{r_{ph}}{\sqrt{1-\frac{r_{ph}^2 \log \left(\frac{6 M \xi }{r_{ph}^3}+1\right)}{3 \xi }}}\label{rsh}.
\end{eqnarray}
On the other hand, emission rate has the following from \cite{Cardoso,Das,emissionWei}
\begin{eqnarray}
\frac{\partial ^2E}{\partial t\partial \omega }=\frac{2 \pi ^2 \omega ^3 \sigma}{e^{\frac{\omega }{T_H}}-1},\label{emission}
\end{eqnarray}
where $\sigma$ represents the cross section and is given by
\begin{eqnarray}
\sigma=\pi r_{sh}^2.\label{sigma}
\end{eqnarray}
By substituting Eqs. \eqref{rsh} and \eqref{sigma}, in Eq. \eqref{emission}, the emission rate could be find. In Fig. \ref{fig:emissionrate}, the demonstration of emission rate for three different $\xi$ is shown.
As can be seen, decreasing parameter $\xi$ lowered the maximum value of emission rate and causes this maximum occur in a smaller value of $r_h$.
\\Lifetime of a black hole could be found as \cite{remnant,evaporation2}
\begin{eqnarray}
\frac{\text{d}M}{\text{d}\tilde{\tau}}=-\tilde{\alpha}\sigma T_H^4,
\end{eqnarray}
where $\tilde{\alpha}$ represents the product of greybody factor and the radiation constant. Here we assume $\tilde{\alpha}=1$.
Thus
\begin{eqnarray}
\tilde{\tau}=\int^{r_{initial}}_{{r_{rem}}} \frac{d M}{d r}\frac{-d r}{\sigma T_H^4}.
\end{eqnarray}
In case of $\xi=0.005$, life time of the black hole becomes $\tilde{\tau}\sim 10^{31}$, by increasing parameter $\xi$ to
 $\xi=0.050$, evaporation time yields $\tilde{\tau}\sim 10^{33}$, and choosing parameter $\xi$ as $\xi=0.200$, evaporation time becomes as order of $\tilde{\tau}\sim 10^{34}$.
\section{Rosen-Morse Potential and Quasi-Normal Modes}\label{Sec8}
\begin{figure}[ht!]
\centering
	\begin{subfigure}[b]{0.5\textwidth}
	\centering
	\includegraphics[scale = 0.48]{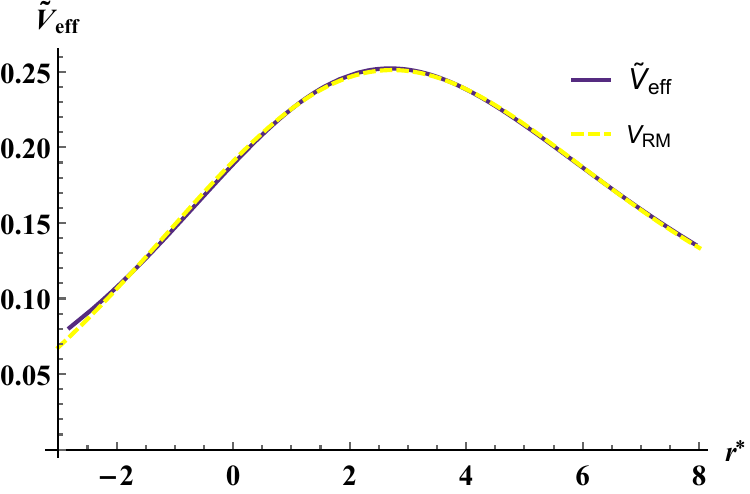} \hspace{-0.2cm}\\
    \end{subfigure}%
    \caption{Fitted $V_{RM}$ to $V$ as a function of $r^*$. Parameters are set as $M=1,\,\xi=0.1,\,l=2,$ and $s=0$.}
    \label{fig:VtildeRM}
\end{figure}

Quasi-normal modes are characteristic oscillations of perturbed black holes, which arise from the interaction of the black hole with its surrounding. These modes provide insights into the astrophysical properties of black holes and have applications in fields like string theory, brane-world models, and the study of quark-gluon plasmas \cite{Konoplya2,Pedrotti}.\\Author in Ref. \cite{Stashko}, reported the quasi-normal modes of metric \eqref{fr} for scalar, vector and Dirac fields using Chebyshev pseudo-spectral and Eikonal methods. In this section, quasi-normal modes of the black hole using Rosen-Morse potential for the expanded form of metric.
\\By expanding logarithm function, metric \eqref{fr} yields to
\begin{align}
\nonumber
\tilde{f}(r)&=1-\frac{2 M}{r}+\frac{6 M^2 \xi }{r^4}\\&-\frac{24 M^3 \xi ^2}{r^7}+\frac{108 M^4 \xi ^3}{r^{10}}-O(\frac{1}{r})^{13}.\label{frExpanded}
\end{align}
The effective potential is obtained from by \cite{Konoplya}
\begin{eqnarray}
V_{eff}(r)=f(r)\left(\frac{1-s^2}{r}\frac{\partial f(r)}{\partial r}+\frac{l (l+1)}{r^2}\right),
\end{eqnarray}
where parameter $l$ represents the angular momentum and $s$ indicates the shape of perturbation. Case $s=0$ represents scalar perturbation and case $s=1$ expresses electromagnetic perturbation.
\\In tortoise coordinate $r^*$ \cite{Konoplya,Heidari}
\begin{align}
r^{*}&=\int\frac{\text{d} r}{f(r)}.
\end{align}
\\Rosen-Morse function \cite{Rosen,Heidari2}
\begin{eqnarray}
V_{RM}=\frac{V_0}{\cosh ^2\alpha  (r^{*}-\bar{r}^{*})}+V_1 \tanh \alpha  (r^{*}-\bar{r}^{*}).
\end{eqnarray}
Here we chose metric $\tilde{f}(r)$ of form Eq. \eqref{frExpanded}.
By fitting potential of $V_{RM}$ to the $V_{eff}(r^{*})$ curve, parameters of RM potential $\alpha,\, V_0,\, V_1,$ and $\bar{r}^{*}$ can be find. Fig. \ref{fig:VtildeRM} shows the fitted Rosen-Morse potential (dashed yellow curve)to the effective potential (purple curve) for $M=1,\,\xi=0.1,\,l=2,$ and $s=0$.
\\Quasi-normal of Rosen-Morse potential are obtained from \cite{Heidari2}
\begin{align}
\frac{\sqrt{\omega ^2-V_1}+\sqrt{V_1+\omega ^2}}{2}=i \alpha  \left(n+\frac{1}{2}\right)\pm\alpha  \sqrt{\frac{V_0}{\alpha ^2}-\frac{1}{4}}\label{RMQNM}.
\end{align}
In this case we used to approximation, the first one in expanding logarithm function, and the second one in fitting potential curve. However the mean absolute error of fitting in our work, for case $l=1$, is as order of $10^{-4}$. The quasi-normal modes of metric \eqref{frExpanded}, using Rosen-Morse approach for cases $\xi=0.1$ and $\xi=0.3$ are reported in table \ref{Table:QNMs}.
\begin{table}[htbp]
		\caption{Quasi-normal modes of fitted Rosen-Morse potential. $M=1,$ and $s=0$.}
		\centering
		\label{Table:QNMs}
		\begin{tabular}{|cc|c|c|}
		\hline
		 $l$ & $n$ & QNMs $(\xi=0.1)$ & QNMs $(\xi=0.3)$\\
\cline{1-4}
		 $1$ & $0$ & $0.301033\, -0.095335 i$ & $0.310015\, -0.087782 i$\\
\cline{2-4}
		 & $1$ & $0.299422\, -0.289816 i$ & $0.309294\, -0.264740 i$\\
		 \hline
		 $2$ & $0$ & $0.491777\, -0.092668 i$ & $0.505708\, -0.085217 i$\\
\cline{2-4}
		 & $1$ & $0.488963\, -0.281114 i$ & $0.504613\, -0.256689 i$\\
\cline{2-4}
		 & $2$ & $0.488090\, -0.471362 i$ & $0.504105\, -0.428995 i$\\
		 \hline
		 $3$ & $0$ & $0.684475\, -0.091866 i$ & $0.703427\, -0.084432 i$\\
\cline{2-4}
		 & $1$ & $0.681599\, -0.277710 i$ & $0.702374\, -0.253970 i$\\
\cline{2-4}
		 & $2$ & $0.679447\, -0.466148 i$ & $0.701412\, -0.424493 i$\\
\cline{2-4}
		 & $3$ & $0.678871\, -0.654984 i$ & $0.701016\, -0.595333 i$\\
		 \hline
		\end{tabular}
	\end{table}
\section{Topological Charge}\label{Sec6}
\begin{figure}[ht!]
\centering
	\begin{subfigure}[b]{0.5\textwidth}
	\centering
	\includegraphics[scale = 0.48]{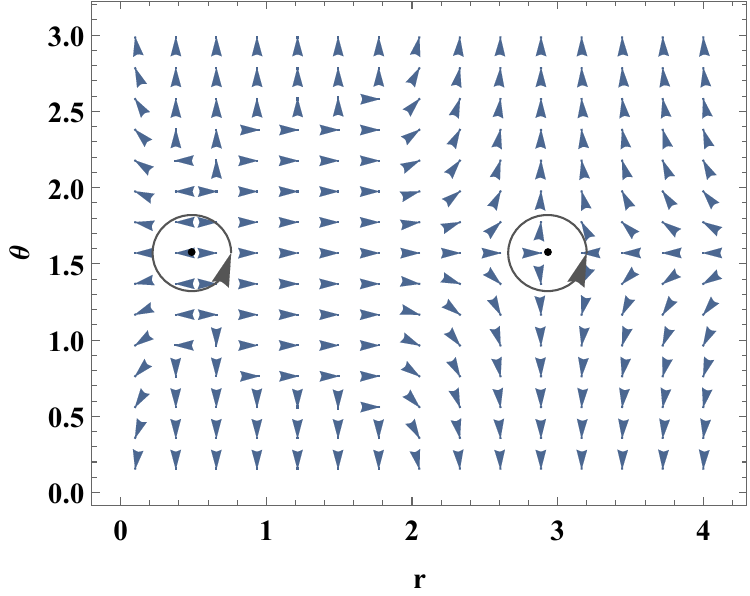} \hspace{-0.2cm}\\
    \end{subfigure}%
    \caption{Vector space for $M=1,$ and $\xi=0.1$. One transition occurs at $r=0.4888$ and another one located at $r=2.9301$, and their relevant topological charge are $w=1$ and $w=-1$, respectively.}
    \label{fig:TopoMet}
\end{figure}

In order to investigate the topology of a black hole, one can assume a potential function in polar coordinate related to the metric \eqref{ds2} as \cite{TopoCharge,Cunha}
\begin{eqnarray}
H(r,\theta)=\frac{1}{\sin\theta}\sqrt{\frac{f(r)}{h(r)}}.\label{Hr}
\end{eqnarray}
Using $H(r,\theta)$ potential, a vector field can be defined as \cite{Cunha}
\refstepcounter{equation}
\begin{align}
\phi_{r}^H=\sqrt{f(r)} \frac{\partial}{\partial r} H(r,\theta),\tag{\theequation a}\label{phimeta}
\\\phi_{\theta}^H=\frac{1}{\sqrt{h(r)}}\frac{\partial}{\partial \theta} H(r,\theta)\tag{\theequation b}\label{phimetb},
\end{align}
for $f(r)$ and $h(r)$ which are defined in Eqs. \eqref{fr} and \eqref{hr}, Eqs. \eqref{phimeta} and \eqref{phimetb} yields to
\refstepcounter{equation}
\begin{align}
\phi_r^H&=-\frac{\csc\theta \left(6 M \xi -3 M r^2+r^3\right)}{6 M \xi  r^2+r^5},\tag{\theequation a}\label{phirh}\\
\phi_{\theta}^H&=-\frac{\cot\theta \csc\theta}{r} \sqrt{\frac{1}{r^2}-\frac{\log \left(\frac{6 M \xi }{r^3}+1\right)}{3 \xi }}.\tag{\theequation b}\label{phith}
\end{align}
The normalized form of the above vectors can be found from
\begin{align}\label{unitv}
n_r^i=\frac{\phi_{r}^i}{||\phi||},\;
n_{\theta}^i=\frac{\phi_{\theta}^i}{||\phi||},
\end{align}
where index $i$ indicates the shape of potential field. The presentation of $(n_r,n_\theta)$ vector space is used to illustrate the topological phase transition. The intersection of vectors with opposite directions indicates a phase transition. Each topological phase transition can be assigned a topological charge, which is calculated in the following.
\\Deflection angle of the topological charge is given by \cite{TopoCharge,TopoDr}
\begin{eqnarray}
\Omega=\arctan \frac{\phi_\theta}{\phi_r}=\arctan\frac{n_\theta}{n_r},
\end{eqnarray}
thus
\begin{eqnarray}
d \Omega=\frac{n_r \partial n_{\theta}-n_{\theta} \partial n_r}{n_{r}^2+n_{\theta}^{2}}\label{domega},
\end{eqnarray}
Using Eqs. (\ref{Hr}) and (\ref{domega}), one can see that $d \Omega$ has one or more poles which are located at $\theta = \pi/2$, and $r_p$ which is the  solution(s) of Eq. (\ref{rph}).
\\In order to release the singularity, the following variables are used \cite{TopoCharge}
\begin{align}
\nonumber
r &= a \cos\vartheta +r_{p},
\\\theta &= b \sin\vartheta+\frac{\pi}{2}
\end{align}
Thus the topological charge, called winding number, is obtained from
\begin{eqnarray}
w_i=\frac{\Delta\Omega}{2\pi}=\frac{1}{2\pi}\oint_C n_r \partial_\vartheta n_\theta - n_\theta \partial_\vartheta n_r.
\end{eqnarray}
Parameter $w_i$ takes $0,\,\pm 1$, that the values $\pm 1$ indicate the phase transition, and the value $0$ indicates no phase change in the black hole. The whole charge $W$ is calculated from the summation of all $w_i$s.
\\Fig. \ref{fig:TopoMet} demonstrates the vector space of $(n_r^H,\,n_{\theta}^H)$, when parameters are set as $M=1,$ and $\xi =0.1$.
As evident, two topological charge existed and the whole charge is $W=0$.
\section{Temperature and Free Energy Phase Transitions}\label{Sec7}
\begin{figure}[ht!]
\centering
	\begin{subfigure}[b]{0.5\textwidth}
	\centering
	\includegraphics[scale = 0.48]{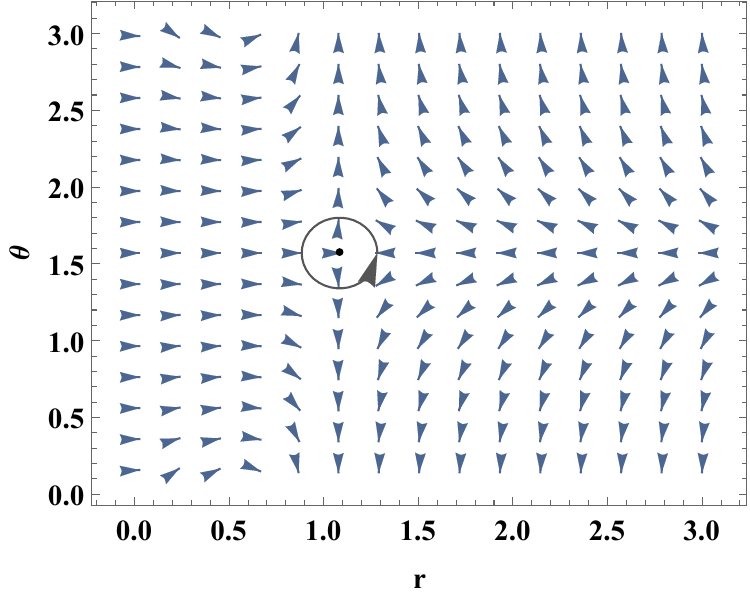} \hspace{-0.2cm}\\
    \end{subfigure}%
\hfill
\caption{Temperature topological charge for $\xi=0.1$. Phase transition occurs at $r_h=1.0837$, winding number is $w=-1$.}
\label{fig:PhaseTemp}
\end{figure}

\begin{figure}[ht!]
\centering
	\begin{subfigure}[b]{0.5\textwidth}
	\centering
	\includegraphics[scale = 0.48]{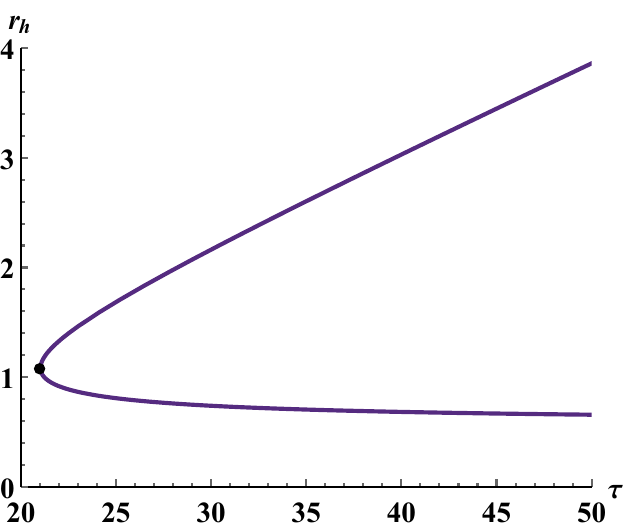} \hspace{-0.2cm}\\
    \end{subfigure}%
\hspace{-0.2cm}
\caption{$r_h$ versus $\tau$, in case of $\xi =0.1$. critical point is located at $(21.0328,\,1.0837)$, and the topological charge of black hole is zero across other regions.}
\label{fig:tau}
\end{figure}

\begin{figure}[ht!]
\centering
	\begin{subfigure}[a]{0.5\textwidth}
	\centering
	\includegraphics[scale = 0.48]{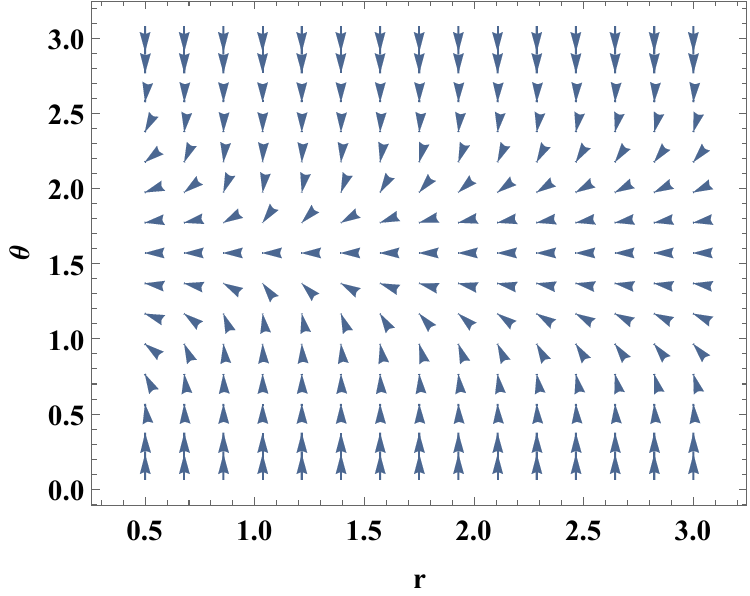} \hspace{-0.2cm}\\
	\caption{Free Energy where $\tau<\tau_c$.}\label{fig:tau20}
    \end{subfigure}%
\hspace{-0.2cm}
\hfill
	\begin{subfigure}[b]{0.5\textwidth}
	\centering
	\includegraphics[scale = 0.48]{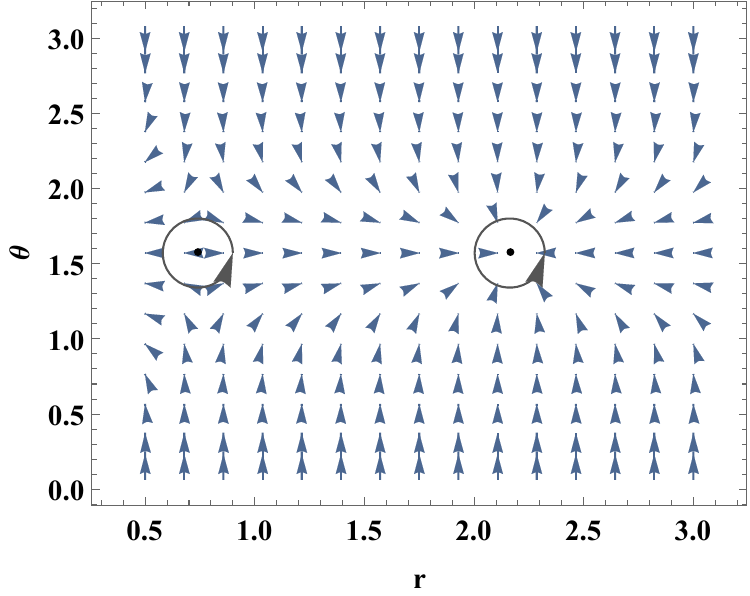} \hspace{-0.2cm}\\
	\caption{Free Energy where $\tau>\tau_c$.}\label{fig:tau30}
    \end{subfigure}%
\hfill
\caption{Free energy topological charge for $\xi=0.1$, and $\tau=30$. At $r_h=0.7401, w=1$, and $r_h=2.1624, w=-1$. In this case the whole topological charge is $W=0$.}
\label{fig:PhaseHelm}
\end{figure}

Another field can be written in the following form \cite{TopoTempWei, TopoDr3}
\begin{eqnarray}
\Phi =\frac{1}{\sin\theta} T_H\label{THfield},
\end{eqnarray}
that $T_H$ indicates the Hawking temperature. Relevant vectors of the field (\ref{THfield}) are defined as
\refstepcounter{equation}
\begin{align}
\phi_{r_h}^{T_H}&=\frac{\partial}{\partial r_h}\Phi\tag{\theequation a},\\\phi_{\theta}^{T_H}&=\frac{\partial}{\partial \theta}\Phi\tag{\theequation b},
\end{align}
by using Eq. \ref{TH}, vectors yield to
\refstepcounter{equation}
\begin{align}
\phi_{r_h}^{T_H}&=\frac{\csc\theta}{4 \pi  \xi r_h^2} \left(2 \xi -e^{-\frac{3 \xi }{r_h^2}} \left(6 \xi +r_h^2\right)+r_h^2\right),\tag{\theequation a}\\
\phi_{\theta}^{T_H}&=-\frac{r_h \cot\theta \csc\theta}{12 \pi  \xi } \left(-\frac{6 \xi }{r_h^2}-3 e^{-\frac{3 \xi }{r_h^2}}+3\right).\tag{\theequation b}
\end{align}
As mentioned in Eq. (\ref{unitv}), the normalized form of this vectors can be write.
\\The illustration of $(n_{r_h}^{T_H},\,n_{\theta}^{T_H})$ for the case $\xi=0.1$, is shown in Fig. \ref{fig:PhaseTemp}.
\\
Obviously, one topological charge exists in this field and the whole charge is $W=-1$.
\\Free energy of the black hole is obtained in Eq. (\ref{free}). Using the definition $\tau=1/T_H$, Eq. (\ref{free}) is rewritten as \cite{TopoEnergyWei,TopoDr2}
\begin{align}
\nonumber
\mathcal{F}&=M(r_h)- \frac{S}{\tau}\\&=\frac{18 \pi  \xi ^2 \text{Ei}\left(\frac{3 \xi }{r_h^2}\right)+r_h^2 \left(e^{\frac{3 \xi }{r_h^2}} (r_h \tau -6 \pi  \xi )-r_h \tau \right)}{6 \xi  \tau }.\label{FreeE}
\end{align}
Considering $\mathcal{F}$ as another field potential, vector space of this field are given by
\refstepcounter{equation}
\begin{align}
\phi_{r_h}^{\mathcal{F}}&=\frac{\partial \mathcal{F}}{\partial r_h} \tag{\theequation a},\label{vectorHelma}
\\
\phi_{\theta}^{\mathcal{F}}&=-\cot\theta\csc\theta\tag{\theequation b}\label{vectorHelmb}.
\end{align}
Applying $\mathcal{F}$ from Eq. (\ref{FreeE}), to  (\ref{vectorHelma}), vectors of the relevant field are rewritten as
\begin{align}
\phi_{r_h}^{\mathcal{F}}=e^{\frac{3 \xi }{r_h^2}} \frac{\tau  \left(r_h^2-2 \xi \right)-4 \pi  \xi  r_h}{2 \xi  \tau }-\frac{r_h^2 \tau }{2 \xi  \tau }.
\end{align}
At $\phi_{r_h}^{\mathcal{F}}=0$, $r_h$ can be find as a function of $\tau$. The illustration of $r_h-\tau$ where parameter $\xi$ is chose as $\xi=0.1$, is shown in Fig. \ref{fig:tau}.
\\As obvious in Fig. \ref{fig:tau}, in case of $\xi=0.1$, the critical value of $\tau$ is located at $\tau_c =21.0328$, and we expect different behavior of Helmholtz free energy field on both side of $\tau_c$. Thus, vector space of this field in case of $\xi=0.1$ in both regime of $\tau<\tau_c$, and $\tau>\tau_c$ is shown in Figs. \ref{fig:tau20} and \ref{fig:tau30}, respectively.
As expected, in $\tau>\tau_c$, free energy potential field phases causes two topological charge, while in case of $\tau<\tau_c$, there is no topological charge. However in both cases the whole charge is $W=0$.
\section{Conclusion}\label{Sec10}
In this work, we have investigated some properties of regular black holes. In the first step, by examining the Hawking temperature of the metric $f(r)$, we identified that an increase in the parameter $\xi$ leads to an increase in the temperature of the black hole. We also calculated the remnant radius and remnant mass of the black hole. In the further investigation of the black hole, we saw how changes in the parameter $\xi$ affect the entropy, heat capacity, and free energy of the black hole. We studied the emission rate and the evaporation process of the black hole for different values of $\xi$. We then expanded the metric and calculated the quasi-normal modes for this metric and observed the effect of changing the parameter $\xi$. In the final step, we investigated the topological behavior of the black hole due to different vector fields and calculated the topological charge for the relevant fields.
\hspace{4cm}
\section*{Acknowledgements}
The research was partially supported by the Long-Term Conceptual Development of a University of Hradec Králové for 2023, issued by the Ministry of Education, Youth, and Sports of the Czech Republic.
\nocite{*}

\bibliographystyle{plain}

\begin{thebibliography}{}

\end{thebibliography}


\begin{thebibliography}{99}
\section*{References}
\bibitem{Clifton}
T. Clifton, P. G. Ferreira, A. Padilla and C. Skordis, Modified gravity and cosmology, Phys. Rept. 513, 1 (2012).

\bibitem{Shankaranarayanan}
S. Shankaranarayanan, J. P. Johnson, Modified theories of gravity: Why, how and what?, Gen. Relativ. Gravit. 54, 44 (2022).

\bibitem{Capozziello}
S. Capozziello and M. De Laurentis, Extended theories of gravity, Phys. Rept. 509, 167 (2011).

\bibitem{Bonanno}
A. Bonanno, D. Malafarina, and A. Panassiti, Dust Collapse in Asymptotic Safety: A Path to Regular Black Holes, Phys. Rev. Lett. 132, 031401 (2024).

\bibitem{Nojiri}
S. Nojiri and S. D. Odintsov, Unified cosmic history in modified gravity: from F(R) theory to Lorentz non-invariant models, Phys. Rept. 505, 59 (2011).

\bibitem{Ashtekar}
A. Ashtekar and M. Bojowald, Quantum geometry and the Schwarzschild singularity, Class. Quantum Gravity, 23, 391 (2006).

\bibitem{Stashko}
O. Stashko, Quasinormal modes and gray-body factors of regular black holes in asymptotically safe gravity, arXiv: 2407.07892 (2024).

\bibitem{remnant}
A. A. Ara\'ujo Filho, S. Zare, P. J. Porfírio, J. Kříž, H.Hassanabadi, Thermodynamics and evaporation of a modified schwarzschild black hole in a non–commutative gauge theory, Phys. Lett. B 838, 137744 (2023).

\bibitem{remnant2}
S. Zare, L.M. Nieto, F. Hosseinifar, X.-H. Feng, H. Hassanabadi, Influences of modified Chaplygin dark fluid around a black hole, arXiv: 2407.12142 (2024).

\bibitem{Claudel}
C. M. Claudel, K. S. Virbhadra and G. F. R. Ellis, The geometry of photon surfaces,  J. Math. Phys. 42, 818 (2001).

\bibitem{Virbhadra}
K. S. Virbhadra, G. F. R. Ellis, Gravitational lensing by naked singularities, Phys. Rev. D 65, 103004 (2002).

\bibitem{Das}
Sumit R. Das and Samir D. Mathur, Comparing decay rates for black holes and D-branes, Nucl. Phys. B 478, 561 (1996),

\bibitem{Cardoso}
V. Cardoso, M. Cavaglia, and L. Gualtieri, Black hole particle emission in higher-dimensional spacetimes, Phys. Rev. Lett. 96, 071301 (2006).

\bibitem{emissionWei}
S.-W. Wei, Y.-X. Liu, Observing the shadow of Einstein-Maxwell-Dilaton-Axion black hole, JCAP 11, 063 (2013).

\bibitem{evaporation2}
N. Heidari, J. A. A. S. Reis, H. Hassanabadi, et al., The impact of an antisymmetric tensor on charged black holes: evaporation process, geodesics, deflection
angle, scattering effects and quasinormal modes, arXiv: 2404.10721 (2024).

\bibitem{Konoplya2}
R. A. Konoplya and A. Zhidenko, Quasinormal modes of black holes: From astrophysics to string theory, Rev. Mod. Phys. 83, 793 (2011)

\bibitem{Pedrotti}
D. Pedrotti, and S. Vagnozzi, See the lightning, hear the thunder: quasinormal modes-shadow correspondence for rotating regular black holes, arXiv: 2404.07589 (2024).

\bibitem{Konoplya}
 R.A. Konoplya, A.F. Zinhailo, J. Kunz, Z. Stuchlik, A. Zhidenko, Quasinormal ringing of regular black holes in asymptotically safe gravity: the importance of overtones, JCAP 10, 091 (2022).

\bibitem{Heidari}
 N. Heidari, H. Hassanabadi, A. A. Ara\'ujo Filho, andJ. Kriz, Exploring non-commutativity as a perturbationin the schwarzschild black hole: quasinormal modes, scattering, and shadows, Eur. Phys. J. C 84, 566 (2024).

\bibitem{Rosen}
N. Rosen and P. M. Morse, On the vibrations of polyatomic molecules, Phys. Rev. 42, 210 (1932).

\bibitem{Heidari2}
N. Heidari, H. Hassanabadi, Investigation of the quasinormal modes of a Schwarzschild black hole by a new generalized approach, Phys. Lett. B 839, 137814 (2023).

\bibitem{Cunha}
P. V. P. Cunha and C. A. R. Herdeiro, Stationary Black Holes and Light Rings, Phys. Rev. Lett. 124, 181101 (2020).

\bibitem{TopoCharge}
S.-W. Wei, Topological charge and black hole photon spheres, Phys. Rev. D 102, 064039 (2020).

\bibitem{TopoDr}
H. Chen, M.-Y. Zhang, H. Hassanabadi, Z.-W. Long, Thermodynamic topology of Phantom AdS Black Holes in Massive Gravity, arXiv: 2404.08243 (2024).

\bibitem{TopoTempWei}
S.-W. Wei and Y-X. Liu, Topology of black hole thermodynamics, Phys. Rev. D 105, 104003 (2022).

\bibitem{TopoDr3}
M.-Y. Zhang, H. Chen, H. Hassanabadi, Z.-W. Long and H. Yang, Topology of nonlinearly charged black hole chemistry via massive gravity, Eur. Phys. J. C 83, 773 (2023).

\bibitem{TopoEnergyWei}
S.-W. Wei, Y.-X. Liu, and R. B. Mann, Black hole solutions as topological thermodynamic defects, Phys. Rev. Lett. 129, 191101 (2022).

\bibitem{TopoDr2}
M.-Y. Zhang, H. Chen, H. Hassanabadi, Z.-W. Long and H. Yang, Thermodynamic topology of Kerr-Sen black holes via Rényi statistics, Phys. Lett. B 856, 138885 (2024).
\end{thebibliography}

\end{document}